\documentclass[reprint,aps,prl]{revtex4-1}

\usepackage{graphicx}
\usepackage{amssymb}
\usepackage{epstopdf}
\usepackage{graphicx}
\usepackage{epstopdf}
\usepackage{amsmath}
\usepackage{color}
\usepackage{xspace}
\usepackage{natbib}

\newcommand\eg{\textit{e.g.}\xspace}
\newcommand\ie{\textit{i.e.}\xspace}

\renewcommand\Im{\mathrm{Im}}

\def \DS {\sigma}

\def \kpar {k_{\parallel}}
\def \kperp {k_{\perp}}

\def \Lpar {L_{\parallel}}
\def \Reff {R_{\mathrm{c}}}

\def \vth {v_{\scriptsize{\mathrm{th}}}}
\def \vth {v_{\scriptsize{\mathrm{T}}}}

\def \kp {k_\alpha^{\mathrm{p}}}
\def \ks {k_\psi^{\mathrm{s}}}
\def \kt {k_\alpha^{\mathrm{t}}}

\newcommand{\vperp}{\ensuremath{v_{\perp}}}

\providecommand\bnabla{\boldsymbol{\nabla}}

\def \kyo {k_{y}^{\mathrm{o}}}
\def \lyo {\ell_{y}^{\mathrm{o}}}

\def \oNL {\omega_{\mathrm{NL}}}

\def \gs {\gamma_s}
\def \gz {\gamma_z}

\newcommand{\zon}[1]{\left<#1\right>_{\psi}}

\begin{document}

\title{Distinct turbulence saturation regimes in stellarators}
\author{G. G. Plunk}
\email{gplunk@ipp.mpg.de}
\affiliation{Max-Planck-Institut f\"ur Plasmaphysik, Wendelsteinstr. 1, 17491 Greifswald, Germany}
\author{P. Xanthopoulos}
\affiliation{Max-Planck-Institut f\"ur Plasmaphysik, Wendelsteinstr. 1, 17491 Greifswald, Germany}
\author{P. Helander}
\affiliation{Max-Planck-Institut f\"ur Plasmaphysik, Wendelsteinstr. 1, 17491 Greifswald, Germany}

\begin{abstract}
In the complex 3D magnetic fields of stellarators, ion-temperature-gradient turbulence is shown to have two distinct saturation regimes, as revealed by petascale numerical simulations, and explained by a simple turbulence theory.  The first regime is marked by strong zonal flows, and matches previous observations in tokamaks. The newly observed second regime, in contrast, exhibits small-scale quasi-two-dimensional turbulence, negligible zonal flows, and, surprisingly, a weaker heat flux scaling. Our findings suggest that key details of the magnetic geometry control turbulence in stellarators.
\end{abstract}

\maketitle

{\bf Introduction.}  The success of magnetic confinement fusion experiments depends on controlling heat loss from the plasma contained within them.  In the case of optimized stellarators, particularly at the outer radial locations of the plasma, this loss is thought to be mainly caused by transport due to micro-turbulence, which exists on scales much smaller than the device size.  A thorough understanding of the fundamental nature of the turbulence is thus strongly desired.  First and foremost, it would be invaluable to identify generic turbulence scaling laws, which can arise from nonlinear saturation processes that are common between different devices.  Indeed, this could greatly facilitate the understanding and interpretation of observations from experiments and numerical simulations, especially when exploring new devices or parameter regimes.

However, given the variety and inherent complexity \cite{boozer-rmp} of stellarators, it is reasonable to ask whether such laws could be expected to apply.  At first glance, the prospect may already seem unrealistic in the simpler context of the tokamak.  That is, although dimensional analysis theoretically constrains transport fluxes to follow gyro-Bohm scaling \cite{hagan-frieman}, these fluxes depend in principle on a host of dimensionless system parameters, including ratios derived from all the characteristic macroscopic scales.   Furthermore, in topologically toroidal magnetic geometry, there is no single scale corresponding to the continuous spatial dependence of the magnetic field, and thus it seems necessary to include detailed geometric information in any theory that hopes to predict the properties of the turbulence.

Despite these apparent challenges, certain generic scaling law behavior in the properties of ion temperature gradient (ITG) turbulence was indeed observed in tokamak simulations by \citet{barnes-cb-itg}.  These laws involved only a small number of characteristic scales, and the magnetic geometry, in particular, was reduced to a single scale, the parallel connection length $\Lpar$, which limits the size of turbulence in the direction parallel to the magnetic guide field.

In this Letter, we present a theoretical study of the turbulence found in two different optimized stellarators, Wendelstein 7-X (W7-X) \cite{w7x} (high-mirror vacuum configuration) and the Helically Symmetric eXperiment (HSX) \cite{hsx}, as compared with that of a conventional circular tokamak. We find qualitatively similar properties in W7-X and the tokamak, but discover a new regime of ITG turbulence in HSX.  Remarkably, this turbulence does not involve significant zonal flow (ZF) activity (flows that regulate turbulence in tokamaks), away from the threshold of turbulence onset.  This allows for the uninhibited growth of small-scale quasi-two-dimensional vortices, resulting in a weaker scaling law for the heat flux, and a steeper fluctuation spectrum.  

Beyond addressing the role of zonal flows in stellarators, these observations broaden the notion of generic ITG turbulence in magnetic fusion plasmas, by demonstrating the possibility of a two-dimensional character to the turbulence, in a fundamentally three-dimensional magnetic field; the competition between two and three dimensional dynamics is a theme that spans the field of magnetized plasma turbulence.

Ostensibly, the W7-X and HSX stellarators have similar characteristics related to ITG turbulence: they have, respectively, large aspect ratios of 11 and 8, toroidal field periods of 5 and 4, and small negative values of global magnetic shear.  Thus, the dimensionless parameters for the ITG mode, derived from local characteristic scales, such as the parallel connection length $\Lpar$ and the ion temperature gradient length $L_T$, should be comparable.  Closer inspection of the magnetic geometry, however, reveals key differences in surface compression and local magnetic shear \cite{localshear}, which measures how much neighboring field lines locally diverge from each other, along field line trajectories (the surface average of the local shear equals the global shear mentioned above).  We argue that these features, characteristic of stellarators, due to their ``twisted'' shape, affect the nonlinear stability of ZFs, and thus determine which turbulence regime is accessible.

Actually, previous numerical evidence suggests that stellarators generally do not rely so strongly on ZFs for saturation, in the sense that the artificial suppression of ZFs causes a relatively modest overall effect on the turbulence as compared with tokamaks  \cite{xanthop-prl-2007}.  We can attribute this phenomenon to the strengthening of local secondary modes that feed on the unstable parallel ion flows present in localized turbulence, and we will invoke these modes to explain the new regime of turbulence observed in HSX.

{\bf Simulations.} We simulate stellarator turbulence using the gyrokinetic code GENE \cite{gene, *gene-web, genetbg,gist}.  For simplicity, we assume electrons to be Boltzmann distributed (with the standard subtraction of the flux-surface-averaged component), as we focus on electrostatic ITG driven turbulence.  We use a flux-tube computational domain \cite{beer}, which is a thin box surrounding a selected magnetic field line for one poloidal turn (the short way) around the torus.  This simplification is advantageous here, as it allows for a direct comparison with our theory, which does not include global effects.  The investigation is limited to a single field line, which in both stellarators passes through the outboard midplane where the cross section is bean shaped.

\begin{figure}
\includegraphics[width=0.95\columnwidth]{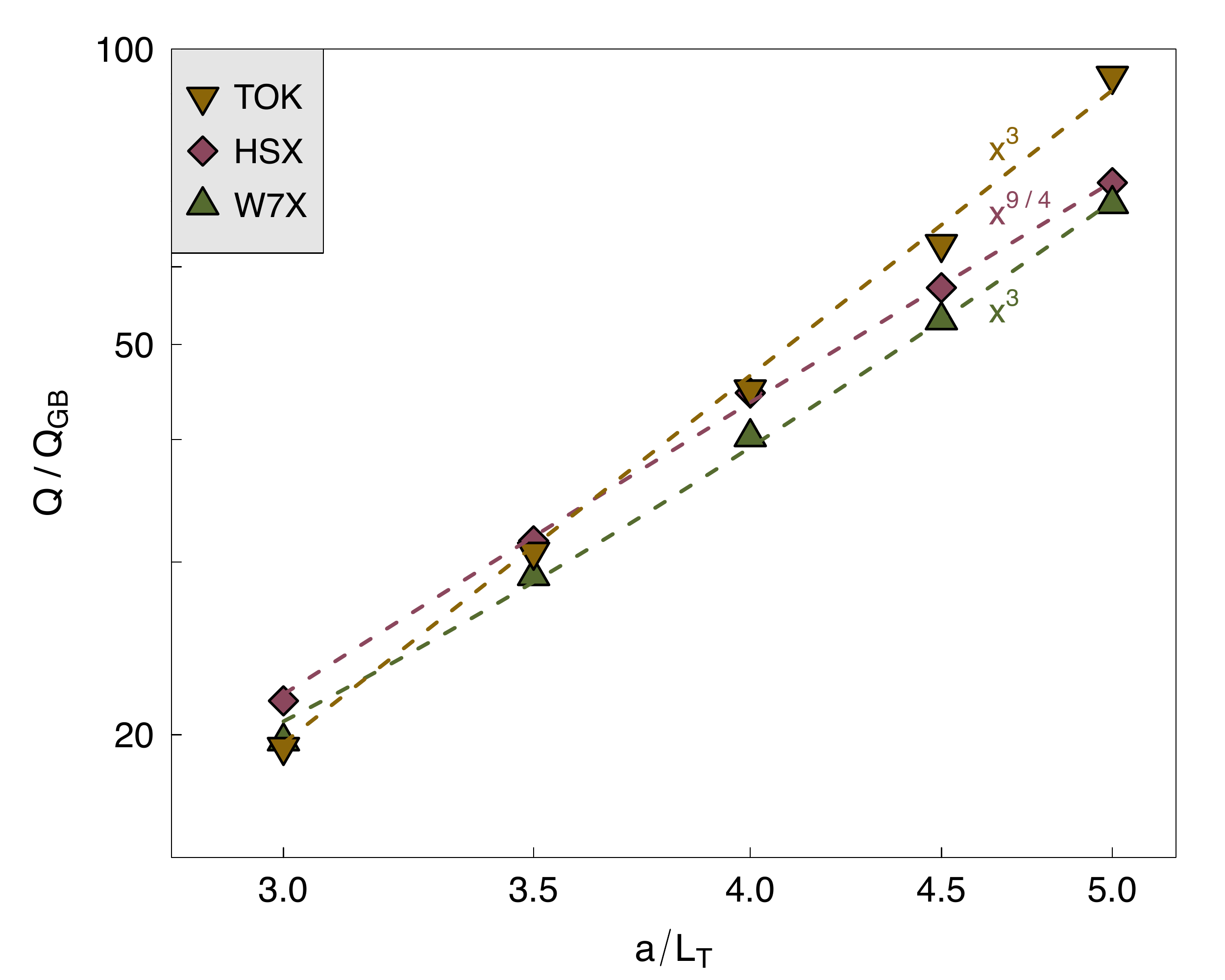}
\caption{Ion heat flux scaling (in gyro-Bohm units) as a function of the ion temperature gradient (normalized to the minor radius), compared with theoretical power laws.}
\label{Q-scaling-fig}
\end{figure}

\begin{figure}
\includegraphics[width=0.95\columnwidth]{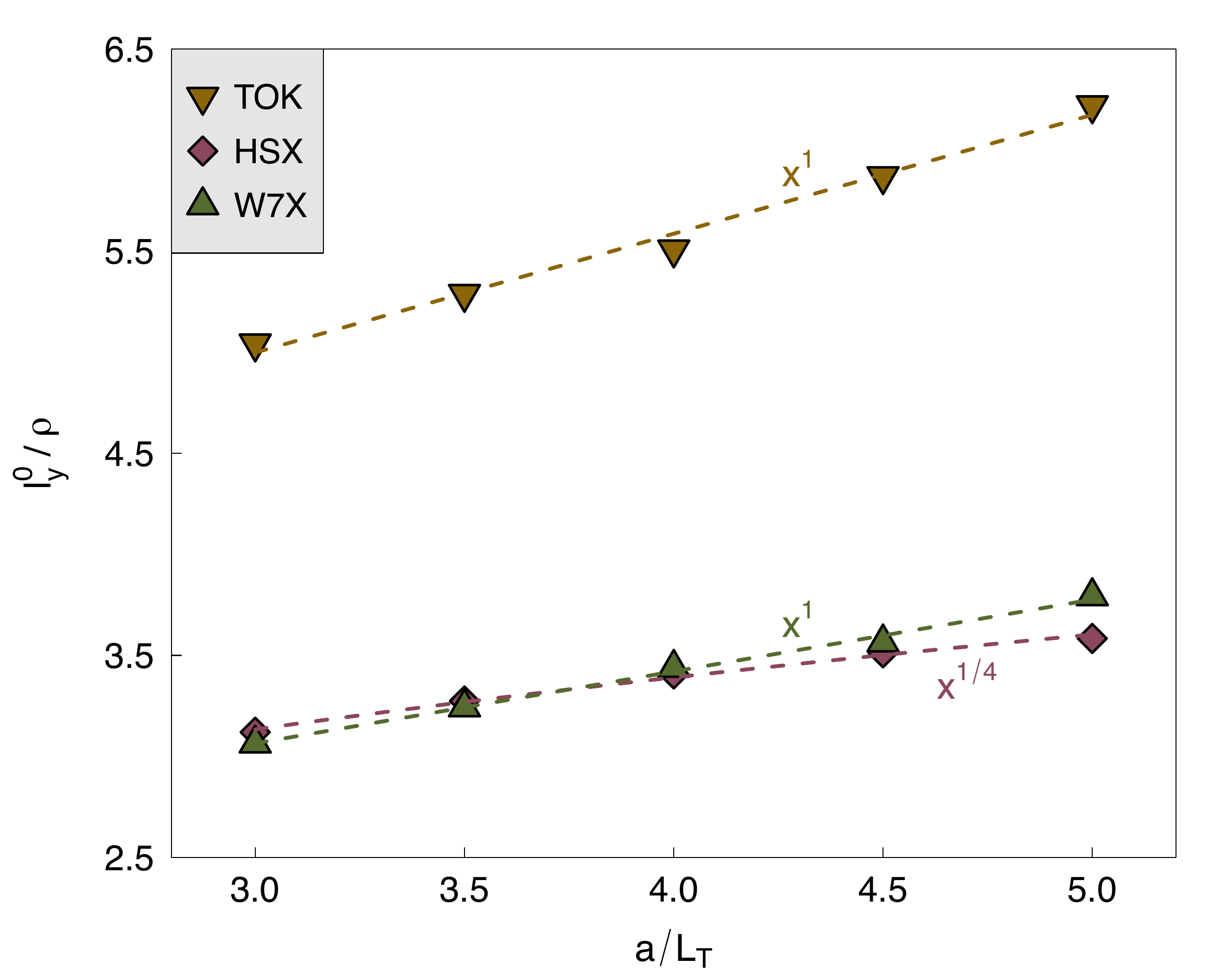}
\caption{Outer scale of the turbulence as measured by the heat flux $\lyo = 1/\kyo = Q^{-1}\sum_{{\bf k}_\perp} k_y^{-1}Q({\bf k}_\perp)$ as a function of the ion temperature gradient.}
\label{ky0-scaling-fig}
\end{figure}

\begin{figure}
\includegraphics[width=0.95\columnwidth]{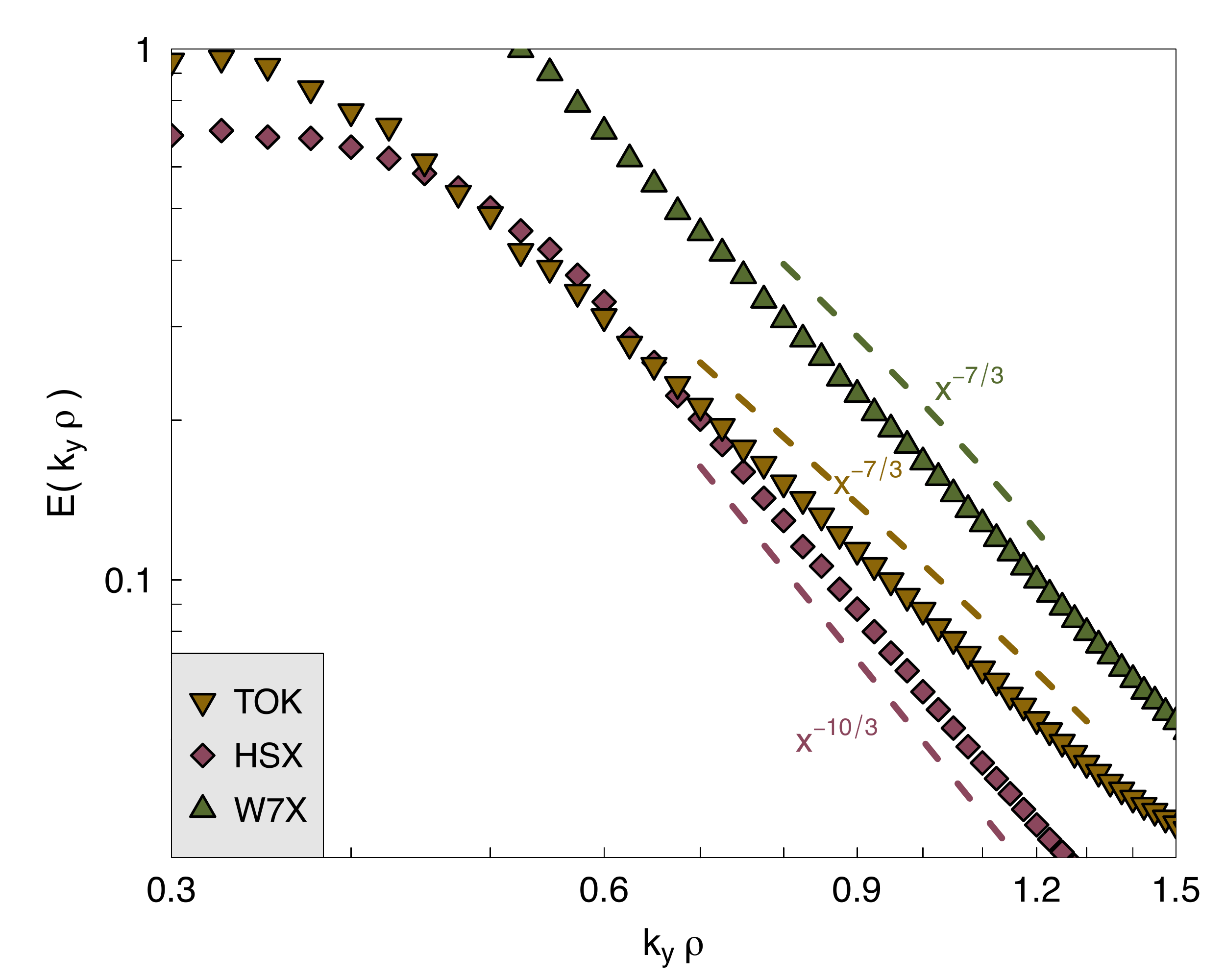}
\caption{Fluctuation spectra compared with theoretical power laws $E(k_y \rho) = \sum_{k_x} |\hat{\phi}|^2/2$.}
\label{spectra-fig}
\end{figure}

In all three devices, we observe a Dimits shift associated with stable ZFs at the weakest temperature gradients, close to the threshold of turbulence onset.  Above the threshold, W7-X has sharply rising transport, matching the theoretical scaling $Q \sim n_0 T_0 \vth \rho^2 \Lpar / L_{T}^3$, known to be satisfied in tokamaks \cite{barnes-cb-itg} (see Fig.~\ref{Q-scaling-fig}); here $n_0$ and $T_0$ are the background ion density and temperature, $L_T^{-1} = -(1/T_0) \, dT_0/dr$, $r$ is the radial coordinate, $\vth = \sqrt{2 T_0/m_i}$ is the ion thermal speed, $m_i$ is the ion mass, and $\rho$ is the ion gyroradius. The turbulence has a characteristic outer scale that depends on the temperature gradient, $\lyo = 1/\kyo \propto L_{T}^{-1}$, as shown in Fig.~\ref{ky0-scaling-fig}.  Such scaling was observed by \citet{barnes-cb-itg}, who associated it with the ``critical balance'' concept; it is also a feature of the slab branch of the ITG mode, since the maximum growth rate is obtained in the resonant limit $\kpar \vth \sim k_y \rho \vth/L_{T}$ of the local dispersion relation \cite{kadomtsev-pogutse}.  The fluctuation spectra in these two cases matches previous observations \cite{barnes-cb-itg}, as shown in Fig.~\ref{spectra-fig}.  

In the case of HSX, however, all of the above measures have different behavior.  The scaling for $Q$ is weaker, as is the scaling for $\lyo$, whereas the fluctuation spectrum turns out to be steeper, following a power law with an exponent of $-10/3$. The scaling of $\lyo$ is a clue that the smaller-scale toroidal branch of the ITG mode drives turbulence in HSX.  Indeed, using the local dispersion relation in the strongly driven limit, the characteristic scale of the toroidal mode may be obtained by balancing the toroidal (magnetic drift) drive with the slab (parallel Landau) drive, \ie by setting the pure toroidal and slab growth rates \cite{biglari, kadomtsev-pogutse} equal, $(\kpar^2 \vth^2 k_y \rho \vth / L_T)^{1/3} \sim k_y\rho \vth/\sqrt{\Reff L_T}$, where $\Reff$ denotes the radius of curvature of the magnetic field.  Setting the parallel wavenumber according to the geometry-dependent connection length, $\kpar \sim 1/\Lpar$, we then obtain $k_y \rho \sim L_T^{1/4}R^{3/4}/\Lpar$.  This $k_y$ is the smallest wavenumber where the toroidal mode can be found, and its dependence on the temperature gradient matches that shown in Fig.~\ref{ky0-scaling-fig}.  Note that this derivation assumes $\kpar \vth \ll \gamma$, \ie the parallel dynamics is so slow, that the modes may be considered as ``quasi-two-dimensional''.  This is also a key distinction with tokamak-type turbulence, driven by the strongly-resonant slab mode, which is fully three dimensional.

The above features exhibited by W7-X and the tokamak are associated with the presence of strong ZFs \cite{plunk-banon-ppcf}, and so it is reasonable to ask whether the difference compared to HSX can be attributed to a difference in ZF activity.  To gain some insight into this question, numerical experiments can be performed where the zonal $E\times B$ flow is artificially set to zero at each time step of the simulation.  In the case of the tokamak, this suppression can result in an increase in transport by two orders of magnitude, or even a failure to saturate altogether.  However for the stellarator devices, the effect is far less dramatic.  Our results show that W7-X has relatively modest change in transport, whereas HSX exhibits no discernible change to the transport level in the final turbulent state.

{\bf Zonal Flow generation.} What causes the strong differences between W7-X and HSX?  Let us first examine the nonlinear ZF drive.  We can estimate ZF growth using secondary instability theory.  To properly account for magnetic geometry, we use magnetic flux coordinates $\psi$ and $\alpha$ such that ${\bf B}_0 = \bnabla \psi \times \bnabla \alpha$.  Denoting the zonal wavenumber $\ks$, assuming $k_\perp^2 \rho_i^2 \ll 1$, and neglecting perpendicular temperature fluctuations and parallel ion motion, the zonal (secondary) growth rate $\gz$ can be expressed in terms of the primary (ITG) mode amplitude $\hat{\phi}_p$ and wavenumber $\kp$ as \cite{plunk-banon-secondary} 

\begin{equation}
\gz = |\kp \ks|\sqrt{2\zon{b_\psi}^{-1}\zon{b_\psi |\hat{\phi}_p|^2}},\label{gamma-z-eqn}
\end{equation}

\noindent where $b_\psi = (\rho \ks |\bnabla \psi|)^2$ and $\zon{.}$ denotes the flux surface average.  Thus, $\gz$ is roughly proportional to the root-mean-squared value of the amplitude $\hat{\phi}_p$ on a flux surface.  To get a sense of how geometry affects ZF growth, we can compare the global result, Eqn.~\ref{gamma-z-eqn}, with what would be obtained from a local theory, at the point where $|\hat{\phi}_p|$ is maximum.  The ratio of the two can be estimated as $\gz/(\kp \ks \mathrm{max}(|\hat{\phi}_p|)) \sim \DS$, where

\begin{equation}
\DS = \sqrt{\zon{|\hat{\phi}_p|^2}}/\mathrm{max}(|\hat{\phi}_p|),
\end{equation}

\noindent with $\mathrm{max}$ denoting the maximum value on the flux surface.  This implies that localized modes (those with small $\DS$) are less effective at driving ZFs than modes that extend across a flux surface. We note that the quantity $\sigma$, as measured in our simulations, is insensitive to the temperature gradient, implying that it is geometry determined.

Turbulence in W7-X is known to be strongly localized \cite{prl14xan}, and indeed we find $\DS = 0.23$ when it is measured in the nonlinear state at the outer-scale wavenumber $\kyo$, whereas the value measured in the HSX stellarator is $\DS = 0.43$, implying a stronger nonlinear drive for ZFs (for comparison, the localization parameter in the tokamak is $\DS = 0.92$). This fact is at odds with the apparent insignificance of ZFs in the final turbulent state.  However, ZFs {\em are} transiently important in the simulations, as evidenced by the appearance of a large burst of heat flux that occurs initially in the simulations where the flows are artificially suppressed.  This seems to indicate that the nonlinear drive mechanism is indeed strong, but that strong damping subsequently arises.

{\bf Zonal flow decay.} We are thus motivated to consider the ZF decay mechanisms, which here include linear collisionless damping (geodesic transfer \cite{scott-geodesic-transfer}), and nonlinear mechanisms such as the tertiary instability \cite{rogers-prl} (we do not include collisions).  To evaluate the importance of geodesic transfer, we perform a series of numerical experiments, with the strength of geodesic curvature artificially modulated.  We find that W7-X is quite sensitive to this procedure, as was previously found \cite{xanthop-geodesic-prl-2011}, while HSX and the tokamak are not.  We note that the transport model of Ref.~\cite{nunami-2013-pop}, which estimates ZF damping using the linear response, might therefore be appropriate for W7-X.  In HSX, however, it seems that ZFs decay by a nonlinear mechanism, \eg the tertiary instability.  A fully three-dimensional theory of the tertiary instability is not analytically tractable, but local theory offers some insight.  As ZF shearing is strongly stabilizing, the tertiary mode is localized to regions of minimal $E \times B$ shearing, and driven by the zonal perpendicular temperature.  The important observation is that this mode depends on finite Larmor radius terms, which are sensitive to magnetic geometry.  To express its growth rate, let us denote the electrostatic potential and perpendicular temperature component of the zonal mode as $\phi_z(\psi)$ and $\chi_z(\psi)$, where $\chi = (q n_0)^{-1}\int d^3{\bf v} (m\vperp^2/4) \delta f$.  The tertiary mode, denoted ($\phi_t$, $\chi_t$), is assumed proportional to $\exp(-i \omega_t t + i \kt \alpha)$, and solved for using gyrokinetic theory, at a fixed location (\eg defined by the toroidal and poloidal angles) on a flux surface $\psi_0$, where $\phi_z''(\psi_0) = 0$.  The result is

\begin{equation}
\omega_t = \omega_E + \frac{\omega_\chi b_\alpha}{2\tau} \pm i\sqrt{-\frac{\omega_\chi^2 b_\alpha^2}{4\tau^2} + \bar{\gamma}_t^2},
\end{equation}

\noindent where $\bar{\gamma}_t = [2 (\rho |\bnabla \psi| \kt)^2 \phi_z'''\chi_z'/\tau]^{1/2}$, $\omega_E = \kt \phi_z'$ is the doppler shift due to the local zonal $E\times B$ velocity, $\omega_\chi = \kt \chi_z'$ is the perturbed diamagnetic frequency, $b_\alpha = (\rho |\bnabla \alpha| \kt)^2$, and $\tau$ is the background ion to electron temperature ratio. The quantity inside the radical, which determines the instability growth rate, depends on magnetic geometry in a subtle way.  To obtain a more transparent form, we can maximize the growth rate over the free parameter $\kt$, and make some simple estimates for the other quantities that appear.  Let us assume that $\phi_z \sim \chi_z$, and that the scale of the zonal mode is comparable to that of the turbulence, so $\phi_z''' \sim \phi_z' (\lyo/|\bnabla \psi|_{\mathrm{ref}})^2$, where the subscript `ref' denotes the reference value measured at the location of peak turbulence intensity.  The zonal flow amplitude $\phi_z'$ may be estimated by the usual balance between the ITG mode growth rate and the zonal shearing rate, $\phi_z' k_\alpha \sim \gamma_L$, thus $\phi_z' \sim |\bnabla \alpha|_{\mathrm{ref}} \rho \vth/L$, where we could take $L = L_T$ or $L = \sqrt{L_T \Reff}$, depending on the drive mechanism.  These estimates yield a maximum tertiary growth rate $\mathrm{max}(\Im[\omega_t]) \sim G_t (\rho/\lyo)^{3/2} \vth/L$, where $G_t$ is a geometric factor, expressed

\begin{equation}
G_t = \frac{|\bnabla \psi|^{3/2}}{|\bnabla \psi|_{\mathrm{ref}}^{3/2}}\frac{|\bnabla \alpha|_{\mathrm{ref}}}{|\bnabla \alpha|}.
\end{equation}

\begin{figure}
\includegraphics[width=0.95\columnwidth]{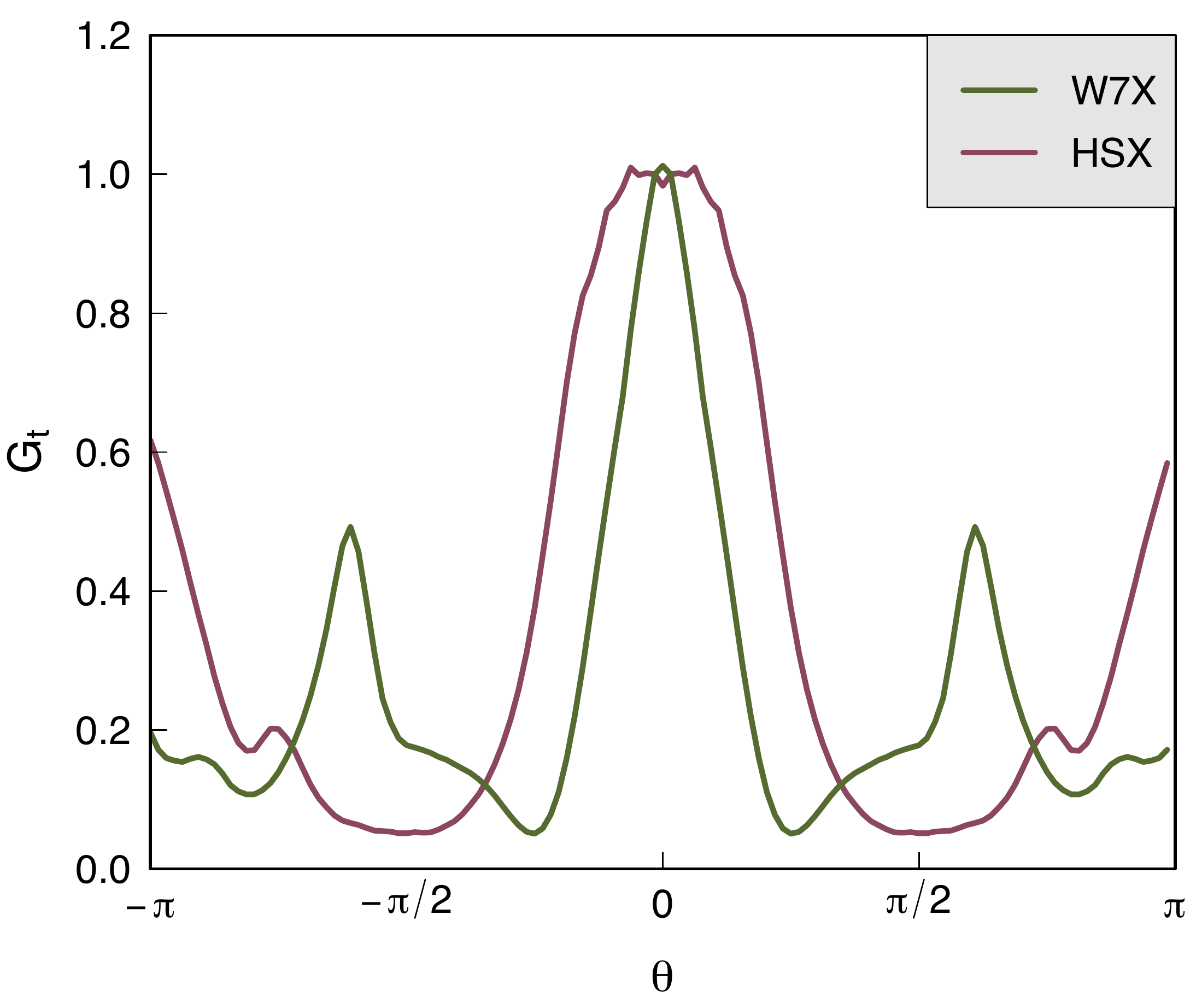}
\caption{Tertiary geometric factor $G_t$ versus the position along the field line, as measured by the poloidal angle $\theta$.}
\label{tertiary-geometric-factor-fig}
\end{figure}

\noindent This factor, which reflects both flux surface compression and local magnetic shear, acts by ``squeezing'' the tertiary mode, \ie limiting its spatial extent, and stabilizing it via the parallel ion Landau resonance. As shown in Fig.~\ref{tertiary-geometric-factor-fig}, the effect is more pronounced in W7-X than HSX, in that the envelope of $G_t$ is narrower. This implies a more stable tertiary mode for W7-X, leading to the observed persistence of ZFs in the system.

{\bf Saturation of toroidal ITG modes.}  As the turbulence observed for HSX is not sensitive to the presence of ZFs, the saturation of ITG modes must be mostly due to eddy/eddy (local) interactions.  We identify the modes driving this turbulence as toroidal ITG modes, due to their small scale and also by the fact that the temperature component grows more strongly than the density component, which is not the case for the resonant slab ITG mode, driving turbulence at large scales.  In the tokamak, these toroidal modes grow to extremely large amplitudes in simulations where ZFs are artificially suppressed.  This behavior has been attributed to the weak variation of the toroidal ITG mode along the field line, and resulting weak parallel ion flow, which makes the mode inherently stable nonlinearly \cite{jenko-dorland-prl}.  The localization of modes by local magnetic shear seems to make stellarators immune to this problem, presumably because local interactions are sufficiently strong to bring about saturation.

Following the above considerations, we propose a model to describe the saturation of toroidal ITG modes by the parallel-ion-flow instability \cite{cowley-kulsrud}.  Let us estimate the linear growth rate of the toroidal ITG mode \cite{biglari} by the (strongly-driven) fluid result $\gamma_L \sim \rho \vth/ (\ell \sqrt{\Reff L_T})$, where $\ell \sim 1/\kperp$ indicates the perpendicular scale.  We then estimate the nonlinear decay rate (inverse turnover time) as $\oNL \sim \gs \sim \delta u_\parallel \rho/\ell^2$, where the parallel ion flow can be expressed as $\delta u_\parallel \sim (q\phi/T)\vth (\omega_\parallel/\gamma_L)$, assuming $\omega_\parallel = \vth/\Lpar < \gamma_L$.  Then the saturation amplitude $\phi_\ell$ may be obtained via $\gamma_L \sim \oNL$.  The non-resonant limit gives also the estimates of the outer scale $\lyo \sim \rho \Lpar/(\Reff^{3/4}L_T^{1/4})$, and the ratio $\delta T \sim q \phi \sqrt{\Reff/L_T}$, which, combined with the saturation amplitude, can be used to evaluate $Q \sim \DS n_0 \phi_{\lyo} \delta T_{\lyo}/(B_0 \lyo)$ as

\begin{equation}
\frac{Q}{Q_{\mathrm{GB}}} \sim\Lpar  \DS \frac{\Reff^{5/4}}{L_T^{9/4}},\label{Q-scaling-HSX-eq}
\end{equation}

\noindent where $Q_{\mathrm{GB}} = n_0 T_0 \vth (\rho/\Reff)^2$. This theoretical prediction compares well with the simulations (see Fig.~\ref{Q-scaling-fig}).  Note the strong dependence on geometry via the factor $\DS\Lpar$.  In fact, $\DS$ sets a geometrical limit on $\Lpar$, since the parallel correlation length cannot exceed the envelope of the turbulence intensity, and thus an overall lower transport could be expected in configurations with more localized turbulence.

{\bf Conclusions.} Using petascale numerical simulations and analytical theory, we have investigated two distinct plasma ITG turbulence regimes in stellarators, with different scaling laws (for the turbulent outer scale and transport), physical drives, and saturation mechanisms.  The first regime, represented by W7-X, resembles tokamak turbulence, in which ZFs play an essential role in saturation, and the slab mode prevails as the drive mechanism. In the second, newly found turbulence regime, represented by HSX, ZFs display a minor effect on transport. 

This stellarator-specific regime is theoretically explained with a model describing the saturation of small-scale toroidal ITG modes by a local secondary instability, leading to a weaker turbulence scaling.  Our work highlights the critical role of magnetic geometry in controlling the turbulence properties, in particular via the affect of local magnetic shear and flux surface compression on turbulence localization and {\em nonlinear} ZF stability.  Stellarator configurations favoring strong turbulence localization, like W7-X, are thus predicted to enjoy more stable ZFs, compared to configurations with less localized turbulence.  Nevertheless, we showed that even in the absence of ZFs, the 3D localization appears sufficient, in contrast to the case in tokamaks, to limit the amplitude of linear ITG modes, making  ``zonal flow free'' saturation possible for stellarator turbulence.

{\bf Acknowledgments.}  The GENE simulations were performed on the Hydra (Germany) and Marconi (Italy) supercomputers.

\bibliography{Stellarator-ITG-scaling-Lett}

\pagebreak

{\bf Erratum.}  In Figs.~\ref{Q-scaling-fig} and \ref{spectra-fig}, dashed curves are provided to compare the simulation results with theoretical power laws.  The curves corresponding to HSX and the tokamak are simple power laws.  However, each of the curves corresponding to W7-X was calculated with a constant offset (\ie of the form $a + b x^r$).  Since offsets are not predicted by the theory, it is incorrect to compare the associated exponents with the theoretical ones.  The data, if fitted instead to pure power laws, would yield exponents between the theoretical values for the tokamak and HSX cases; that is, they would be between $9/4$ and $3$ for Fig.~\ref{Q-scaling-fig}, and between $-7/3$ and $-10/3$ for Fig.~\ref{spectra-fig}.  Thus, we must modify a conclusion of the Letter, namely that W7-X is in the same regime as the tokamak.  Instead, it appears to be an intermediate case.  In support of this conclusion, we note that, comparing the importance of zonal flows in the three cases, W7-X is also observed to be the intermediate case.

We also note that an offset was used in Fig.~\ref{ky0-scaling-fig} to calculate the theoretical fits, but in this case for both W7-X and the tokamak (but not HSX).  A pure power law fit applied to the data yields, in both cases, exponents somewhat below $1/2$.  This does not affect the conclusions of our Letter, but may be relevant to understanding the saturation mechanism that operates in the tokamak regime; we, however, consider this matter to be outside the scope of our work.

The main conclusions of our work, in particular the observation, description, and explanations of the new saturation regime, exemplified by HSX, remain unchanged.

\end{document}